\documentstyle[11pt,paspconf,psfig]{article}

\begin{document}

\newcommand{\uv}{(u,v)}
\newcommand\degd{\ifmmode^{\circ}\!\!\!.\,\else$^{\circ}\!\!\!.\,$\fi}
\newcommand{\etal}{{\it et al.\ }}
\def\la{\mathrel{\mathchoice {\vcenter{\offinterlineskip\halign{\hfil
$\displaystyle##$\hfil\cr<\cr\sim\cr}}}
{\vcenter{\offinterlineskip\halign{\hfil$\textstyle##$\hfil\cr
<\cr\sim\cr}}}
{\vcenter{\offinterlineskip\halign{\hfil$\scriptstyle##$\hfil\cr
<\cr\sim\cr}}}
{\vcenter{\offinterlineskip\halign{\hfil$\scriptscriptstyle##$\hfil\cr
<\cr\sim\cr}}}}}
\def\ga{\mathrel{\mathchoice {\vcenter{\offinterlineskip\halign{\hfil
$\displaystyle##$\hfil\cr>\cr\sim\cr}}}
{\vcenter{\offinterlineskip\halign{\hfil$\textstyle##$\hfil\cr
>\cr\sim\cr}}}
{\vcenter{\offinterlineskip\halign{\hfil$\scriptstyle##$\hfil\cr
>\cr\sim\cr}}}
{\vcenter{\offinterlineskip\halign{\hfil$\scriptscriptstyle##$\hfil\cr
>\cr\sim\cr}}}}}

\title{VLBA Imaging at 7 mm and Linear Polarimetric Observations 
at 6 cm and 3 mm of Sagittarius A*}
\author{Geoffrey C. Bower and Heino Falcke}
\affil{Max Planck Institut f\"ur Radioastronomie, Bonn, Germany}

\author{Don Backer and Melvyn Wright}
\affil{Radio Astronomy Laboratory, UC Berkeley, Berkeley, CA}

\begin{abstract}
We summarize the results of 7 mm VLBA imaging of Sgr A* and
discuss some of the difficulties of accurately constraining the size 
of Sgr A* with VLBI observations.  Our imaging results are fully
consistent with the hypothesis that the VLBA image of Sgr A* is
a resolved elliptical Gaussian caused by the scattering of an 
intervening thermal plasma.  
We show that determination of the minor axis size at 7 mm with the
VLBA is very unreliable.

We also present new polarimetric observations
from the VLA and from the BIMA array of Sgr A*.  At 4.8 GHz, we find
an upper limit to the polarization of 0.1\%.  At 86 GHz, we report a
marginal detection of $1 \pm 1$\% linear polarization.  We discuss
the effects of interstellar propagation on the linear polarization and
consider the significance of very low intrinsic linear polarization
in Sgr A*.
\end{abstract}


\section{7 mm VLBA Imaging of Sgr A*}

\begin{figure}
\caption{Visibility amplitude as a function 
of $uv$ distance for 
Sgr A*.  The solid lines indicate the expectation for a circular
Gaussian with a zero baseline flux of 1.28 Jy and 
FWHM of 0.694, 0.727 and 0.760 mas.  The dashed lines indicate the
expectation for a circular Gaussian with zero baseline fluxes of 1.38
Jy and 1.18 Jy with FWHM of 0.760 and 0.694 mas, respectively.
The noise bias is added in quadrature to the model.
The large triangles indicate the median flux in
a 35 M$\lambda$ bin.  The errors 
indicate the scatter in the data.
\label{fig:sgrauvd}}
\mbox{\psfig{figure=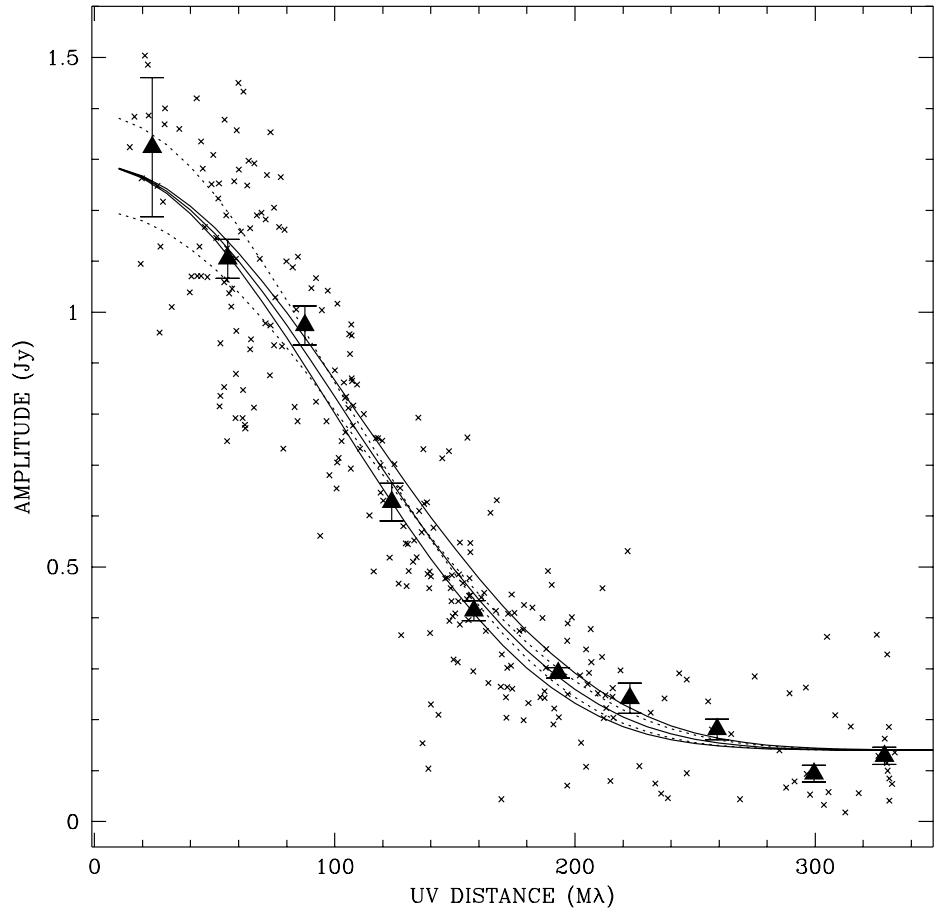,width=1.00\textwidth}}
\end{figure}

We summarize previously-published
results of 7 mm VLBA imaging of Sgr A* (Bower \& Backer 1998).
Observations of Sgr A*
at 43 GHz were performed on 29 September 1994 with the VLBA.
Fringes on Sgr A* were detected to a maximum $\uv$-distance of
250 $M\lambda$.  Fringes to the compact calibrator NRAO~530
were detected on all baselines throughout the experiment.
Imaging and modeling revealed a single elliptical Gaussian component
with a major axis FWHM of $0.762 \pm 0.038$ mas in a position
angle of $77\degd 0 \pm 7\degd 4$.  The axial ratio was measured
to be $0.73 \pm 0.10$.  We show in Figure~\ref{fig:sgrauvd}
the visibility amplitudes
as a function of $\uv$-distance along with model fits.
No secondary structure was detected in the vicinity of
Sgr A* to a limit of 35 mJy.

The measured size is fully consistent with the expected major and
minor axes predicted by lower frequency observation (Yusef-Zadeh et al. 1994,
Lo et al. 1998).  However, as Lo et al. note, this result 
combined with their new result implies a significant deviation
in the axial ratio from that expected due to scattering.  While this is true,
poor North-South resolution and systematic errors in calibration
make this claim uncertain and difficult to prove.

\begin{figure}
\caption{Visibility coverage of detections of Sgr A* at 7 mm by 
Bower \& Backer (1998).  The starred visibilities are those in which
the minor axis dominates the major axis in determination of the visibility
amplitude.
\label{fig:sgrauv}}
\mbox{\psfig{figure=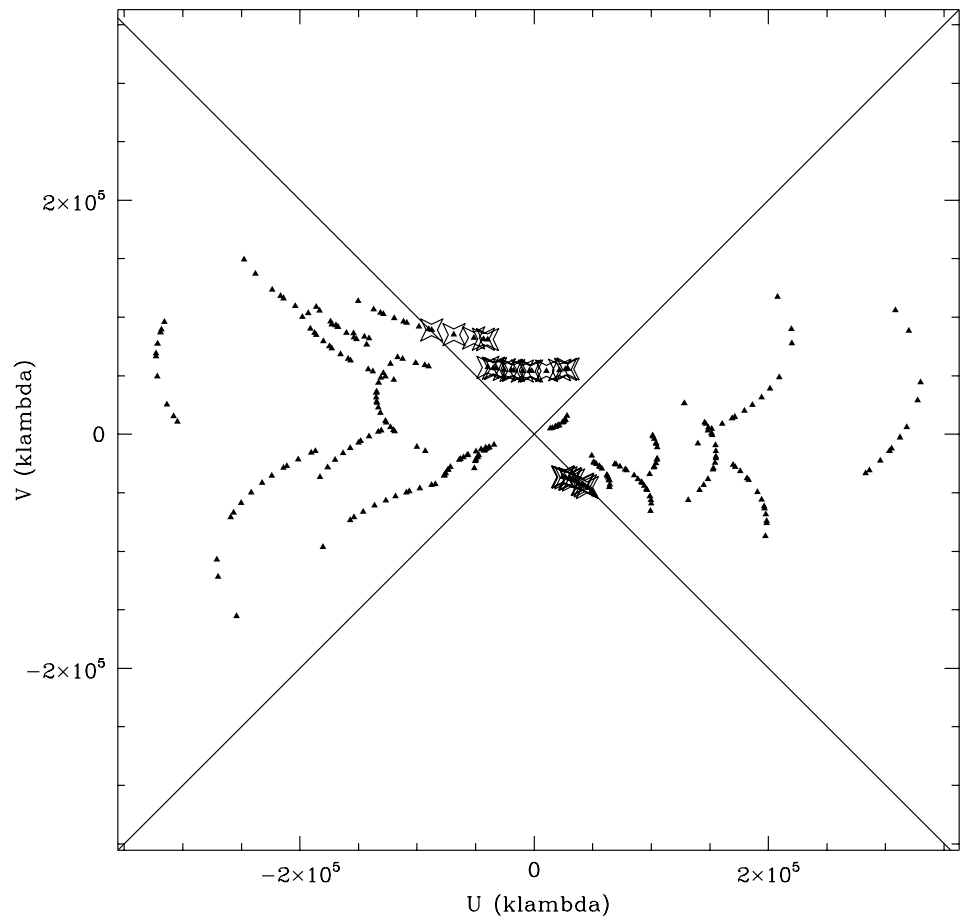,width=1.00\textwidth}}
\end{figure}

{\bf 1) The North-South resolution of the VLBA is inadequate to resolve 
the scattering size at 7mm.}
The extreme southern declination of Sgr A* combined with the 
northern latitudes of the VLBA produces very poor North-South
resolution.  In Figure~\ref{fig:sgrauv}
we show the baselines  on which we made
detections to Sgr A*.  The visibilities that contribute predominantly to
the minor axis size are limited to three baselines, two of which 
involve the Brewster station, one of the VLBA's worst millimeter
sites.  The maximum North-South resolution of
$\sim 100 M\lambda$ on a single baseline is considerably smaller than
the FWHM of the minor axis scattering size $\sim 190 M\lambda$. 
In fact, the size scale of interest --- the difference between 
the scattered and non-scattered minor axis sizes ---
is more than 10 times smaller than the beam size in that direction.

{\bf 2) Observations of Sgr A* with the VLBA are performed at very low
elevation and are difficult to calibrate.}  Variable atmospheric
opacity due to water vapor in the troposphere makes amplitude
calibration particularly difficult at 7 mm.  Atmospheric coherence
times can be very short and variable, as well.  These effects increase
as $\sec z$ with increasing zenith angle, $z$.  Further, {\it a priori}
gain curves at low elevation with VLBA antennas are suspect.  The
Brewster antenna, which provides the two longest North-South
baselines on Sgr A*, never  gets above $13\deg$ elevation.
Hence, it is always looking through at least 4.5 atmospheres.
Transferring gains from a compact calibrator, J1730-13
or J1924-29, for example, can improve amplitude calibration but a 
substantial uncertainty remains due to the large difference in 
contemporaneous elevation.  Amplitude self-calibration is 
not accurate, either, since the density of visibilities in the $\uv$ plane
is low.  

\begin{figure}
\caption{Visibility amplitude corrected for East-West structure against
the North-South spatial frequency $v$.  The lines indicate expected
amplitudes for the mean axial ratio and plus and minus three sigma 
deviations in the axial ratio.  Circled visibilities indicate those on
a baseline with the Brewster station.  These dominate for $v> 60 M\lambda$.
The triangles indicate median amplitudes in
three bins of 40 $M\lambda$ in $v$.
\label{fig:sgrauvddiv}}
\mbox{\psfig{figure=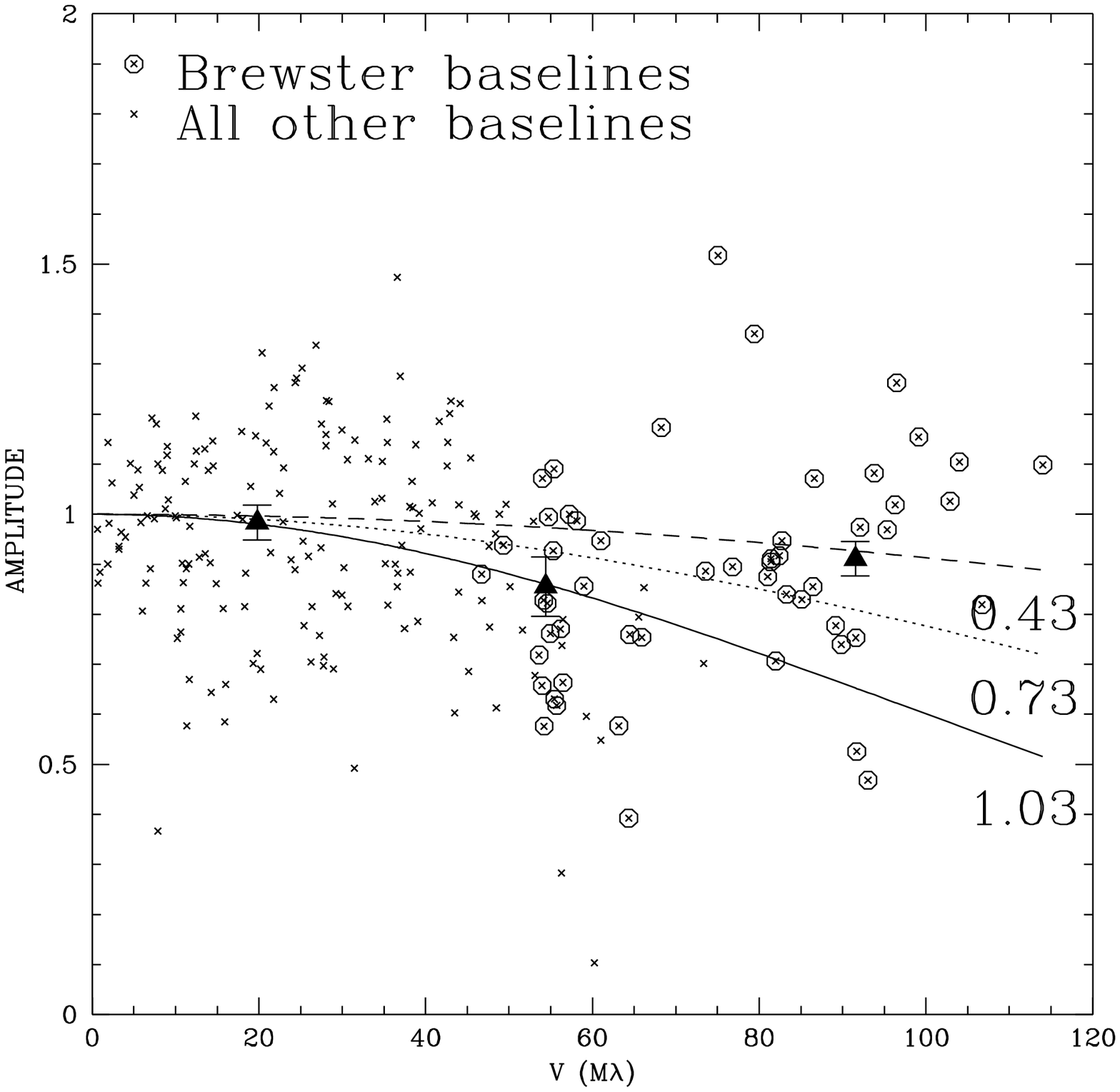,width=1.00\textwidth}}
\end{figure}

One can consider the combined effect of the lack of resolution and
difficulty in amplitude calibration by plotting the amplitudes as a function
of the North-South spatial frequency $v$, after the amplitudes have been
corrected for the East-West structure (Figure~\ref{fig:sgrauvddiv}).
If the North-South axis were unresolved, the residual amplitude would be 
unity.  The considerable scatter in the data indicates the difficulty
in determining the axial ratio accurately.  And, as noted above, 
baselines involving Brewster dominate at large $v$.

What observations can accurately measure the size of Sgr A* at 7 mm?  
The use of the Green Bank Telescope with the VLBA 
will add crucial medium-length North-South baselines to Hancock, Saint Croix,
North Liberty and Fort Davis.  The Large Millimeter Telescope under
construction in Mexico will also add interesting baselines to the 
Southwestern VLBA stations.  Finally, successful adoption of spectral-line
calibration methods using the nearby SiO maser VX Sgr will significantly
improve calibration.

Higher frequency observations are plagued by the same problems of difficult
calibration and poor North-South resolution.  These are discussed in
more detail by Doeleman et al. (1998, these proceedings) and Krichbaum et al. 
(1998, these proceedings).

\section{Linear Polarimetric Observations of Sgr A* at 4.8 and 86 GHz}

Linear polarization stands as one of the few observables of Sgr A*
not extensively investigated observationally or theoretically.  However,
we expect that linear polarization should arise from the cyclo-synchrotron
radiation that is responsible for the radio and millimeter wavelength
spectrum.  In
active galactic nuclei, the degree of linear polarization is a
measure of the order in the magnetic field.  Comparing the evolution
of linear polarization to the evolution of total intensity provides
a strong argument for the existence of shocks in the relativistic
jets of AGN (e.g., Hughes, Aller and Aller 1985).  Detection of 
correlated polarized and total intensity variations in Sgr A* would be
convincing evidence for a jet model.  Uncorrelated variability would
indicate that a different mechanism must be responsible for 
the radio to millimeter spectrum.

\subsection{VLA Observations at 4.8 GHz}

The VLA observed Sgr A* on 10 and 18 April 1998 in the A array at 4.8 GHz
with a bandwidth of 50 MHz.  Instrumental calibration was performed
with the compact sources 1741-038 and 1748-253.  The right-left phase
difference was set with observations of 3C 286.  Baselines longer
than 100 $k\lambda$ were used for Sgr A*.

We summarize the measured polarizations of Sgr A* and several compact
calibrators in Table~\ref{tab:vla6cm}.  
The calibrators GC 441, W56 and W109 belong to the proper motion survey
of Backer and Sramek (1998, these proceedings).
Consistency between the results on the two days indicates the
accuracy of the results.   Polarization was reliably detected from
all sources but Sgr A* and GC 441.  The measured polarization
at the position of
Sgr A* is 0.1\%.  This value is equal to the average off source
fractional polarization in the map and is, therefore, an upper limit.

\begin{table}[htb]
\begin{center}
\caption{Polarized and Total Flux at 4.8 GHz \label{tab:vla6cm}}
\begin{tabular}{llrrrr}
\hline
\hline
Source  & Date & I & P & p & $\chi$ \\
                  &                & (Jy) & (mJy) & (\%) & (deg) \\
\hline
Sgr A*   & 10 Apr & 0.49  &  $<0.63$ & $<0.13$ & \dots \\
         & 18 Apr & 0.47  &  $<0.34$ & $<0.08$ & \dots \\
1741-038 & 10 Apr & 4.90  & 10.6  & 0.22 &  27 \\
         & 18 Apr & 4.94  & 15.3  & 0.31 &  24 \\
1748-253 & 10 Apr & 0.48  &  8.4  & 1.8  &  -15 \\
         & 18 Apr & 0.48  & 11.4  & 2.4  & -16  \\
GC441    & 10 Apr & 0.044 &  $<0.09$ & $<0.20$ &  \dots \\
         & 18 Apr & 0.043 &  $<0.12$ & $<0.28$ &  \dots  \\
W56      & 10 Apr & 0.104 &  2.2  & 2.1 & -66 \\
         & 18 Apr & 0.104 &  2.2  & 2.1 & -66  \\
W109     & 10 Apr & 0.098 &  0.59 & 0.6 & -26 \\
         & 18 Apr & 0.098 &  0.51 & 0.5 & -17  \\
\hline
\end{tabular}
\end{center}
\end{table}

\subsection{BIMA Observations at 86 GHz}

Observations of Sgr A* with the BIMA array were conducted on 10 and 14
March 1998 at 86 GHz with a bandwidth of 800 MHz.  The BIMA array
was configured in its A array with baselines from 20 to 540 $k\lambda$.
Nine antennas were switched systematically between right and left
circular polarization throughout the 5 hour track.  A single 
antenna observed linear polarization constantly.  This permitted 
phase self-calibration for each 10 sec averaging interval.  Amplitude 
calibration was performed with {\it a priori} gain information.
Phase calibration was done with interleaved scans of NRAO~530.  

\begin{figure}[h]
\mbox{\psfig{figure=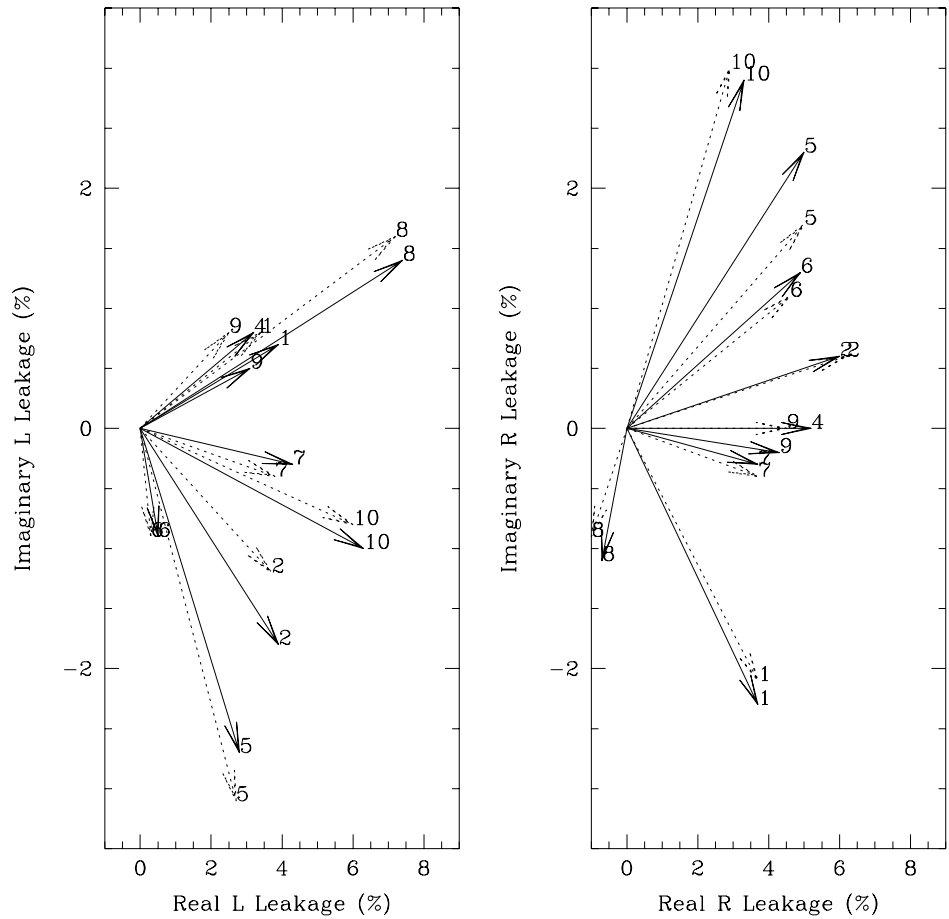,width=1.00\textwidth}}
\caption{D-terms for all antennas determined from SiO maser observations on
25 Feb 98 (solid lines) and 28 Jan 98 (dashed lines).  
The terms differ by 0.3\% on average.
\label{fig:dterm}}
\end{figure}

Instrumental polarization 
calibration was performed in several ways, producing
similar results.  $D$-term solutions were obtained from the 
calibrator NRAO~530 on 14 March, from Sgr A* itself on both days
and from observations of the SiO maser in Orion on 28 January 1998
and 25 February 1998.  A comparison of these $D$-terms shows 
that the instrumental polarization is stable over two months to 
a level of 0.3\% (Figure~\ref{fig:dterm}).

We tabulate the measured polarization on Sgr A* and other sources in
Table~\ref{tab:bima}.  The results for Sgr A* are marginally consistent
with each other and a fractional polarization of $1 \pm 1\% (3\sigma)$.

We note that our observations occurred within days of the peak of the millimeter flare reported
by Tsuboi, Miyazaki and Tsutsumi (1998, these proceedings).  Our mean flux
$S_{86}=1.5 \pm 0.2 {\rm\ Jy}$ is lower than but marginally consistent with
the peak flux of $2.4 \pm 0.4 {\rm\ Jy}$ reported by these authors.  

\begin{table}[htb]
\caption{Polarized and Total Flux at 86 GHz \label{tab:bima}}
\begin{center}
\begin{tabular}{llrrrr}
\hline
\hline
Source  & Date & I & P & p & $\chi$  \\
                  &                & (Jy) & (mJy) &  (\%) & (deg) \\
\hline
Sgr A* & 10 Mar & $ 1.72 \pm 0.075$ &  $22 \pm  6$ & $1.3 \pm 0.3$ & $-41 \pm 8$ \\
       & 14 Mar & $1.37 \pm 0.097$ &  $7 \pm 10$ & $0.5 \pm 0.7$ & $ 16 \pm 41$ \\
3C273  & 10 Mar & $24.7 \pm 1.1$ & $756 \pm 36$ &  $3.1 \pm 0.1$ & $-1 \pm 1$ \\
       & 14 Mar & $23.9 \pm 1.7$ & $802\pm 96$ &  $3.3 \pm 0.3$ & $0 \pm 3$ \\
3C454.3 & 10 Mar & $5.33 \pm 0.37$ & $34 \pm 11$ & $0.6 \pm 0.2$ & $43 \pm 10$ \\
       & 14 Mar & $4.93 \pm 0.13$ & $21 \pm 12$ & $0.4 \pm 0.2$ & $-41 \pm 16$ \\
1733-130 & 14 Mar & $2.82 \pm 0.010$ & $66 \pm 3$ & $2.3 \pm 0.1$ & $-42 \pm 1 $ \\
\hline
\end{tabular}
\end{center}
\end{table}

\subsection{Interstellar Propagation Effects}

The interstellar medium may depolarize a polarized radio wave in two ways:
differential Faraday rotation along the many paths that contribute to
the scatter-broadened image of Sgr A*; and, significant 
rotation of the polarization position angle through the observing
bandwidth.  We consider each of these separately.

The Faraday rotation is
$$\chi_F = {\rm RM}\ \lambda^2,$$
where RM is the rotation measure in rad m$^{-2}$ and $\lambda$ is the
wavelength in meters.  The scattering region will depolarize the signal
if $\delta\chi_F\approx\pi$.  Such variations may occur over multiple
turbulent cells in a scattering screen each of which will have
a different RM.  For $\lambda=6 {\rm\ cm}$, 
$\delta{\rm RM}=900 {\rm\ rad\ m^{-2}}$ (at $\lambda=3 {\rm\ mm}$,
$\delta{\rm RM}=3\times 10^5 {\rm\ rad\ m^{-2}}$).
Over the scattering size of 50 mas at 6 cm,
this corresponds to $\delta{\rm RM}/\delta{\theta}=18000 {\rm\ rad\ m^{-2}
arcsec^{-1}}$.  Observations on the arcsecond to arcminute 
scale of a nonthermal filament within 1 degree of Sgr A* find a maximum
$\delta{\rm RM}/\delta{\theta}=250 {\rm\ rad\ m^{-2} arcsec^{-1}}$
(Yusef-Zadeh, Wardle \& Parastaran 1997).  
Extrapolating their RM structure function to the scattering size implies
$\delta{\rm RM}\approx 50 {\rm\ rad\ m^{-2}}$ 
These values are well below those necessary to 
depolarize Sgr A*.

Do the conditions exist to depolarize Sgr A* in the scattering screen?
The RM is 
$$
{\rm RM} = 0.8 n_e B L {\rm\ rad\ m^{-2}},
$$
where $n_e$ is the electron number density in cm$^{-3}$, 
$B$ is the magnetic field
in $\mu{\rm G}$ and $L$ is the size scale in pc.  
In the thin screen approximation
$L$ must be a fraction of the scattering diameter $\theta_{Sgr A*}$.  
We find
$L \sim 0.1 \theta_{Sgr A*} D_{Sgr A*} \sim 10^{-4} {\rm\ pc}$.
This implies $n_e B \sim 10^7 {\rm\ \mu G\ cm^{-3}}$.  This condition
is matched by that found by Yusef-Zadeh et al. (1994) for the ionized
skins of molecular clouds in the Galactic Center.  However,
the high temperature and low density 
model of Lazio \& Cordes (1998) predicts no depolarization
of Sgr A* for $B < 1{\rm\ G}$.

Finally, depolarization may occur in
the accretion region of Sgr A* where the electron density and
magnetic field strength are large but the length scale is smaller.
In the model of Melia (1994) inside a radius of 0.01 pc, 
$B \ga 100 {\rm\ mG}$ and $n_e \ga 10^4 {\rm\ cm^{-3}}$ for
an accretion rate of $10^{-4} M_{\sun}\ y^{-1}$.  
The ADAF model requires similar conditions (Narayan \etal 1998).
Whether the extreme conditions in the accretion region can cause 
a significant $\delta RM$ depends critically on the scale of 
turbulence.

We now consider the effect of Faraday rotation on a polarized signal measured 
over a specific bandwidth.
Differentiating $\chi_F$
permits us to write the maximum observable rotation measure
as a function of bandwidth $\Delta\nu$ and observing frequency
$${\rm RM_{max}} \approx {1\over 2 } {1 \over \lambda^2}{\nu \over \Delta\nu}.$$
For the observing bandwidths of 50 MHz and 800 MHz at 4.8 GHz and
86 GHz, we find maximum rotation measures of $10^4 {\rm\ rad\ m^{-2}}$
and $10^7 {\rm\ rad\ m^{-2}}$, respectively.  The detection of 
rotation measures in the Galactic Center as large as a few thousand
suggests that the low frequency result may in fact be limited by 
a very high rotation measure (e.g., Yusef-Zadeh, Wardle \& Parastaran, 1997).
We are currently studying spectro-polarimetric observations of the continuum
radiation from Sgr A* at 4.8 and 8.4 GHz in search of linear polarization
with extremely high rotation measures (up to $10^7 {\rm\ rad\ m^{-2}}$).
A preliminary result from this study is that an extreme RM does not
depolarize Sgr A*.

It is not yet certain whether interstellar propagation effects can depolarize
Sgr A*.  Known RM variations are insufficient to depolarize Sgr A*.
However, at least one model for the scattering medium of Sgr A* does
include the necessary conditions.  We now consider the hypothesis of
an intrinsically weakly polarized Sgr A*.

\subsection{Discussion}

The centimeter and millimeter wavelength polarization studies probe
different regions of Sgr A*.  
Spectral studies of Sgr A* (e.g.,
Falcke et al. 1998) reveal that there may be significantly different
sources at centimeter and millimeter wavelengths  which may have
different polarization structures.  
A higher fractional polarization at
86 GHz may be the result of observing at lower opacity and 
over a smaller, more homogeneous region.  
The polarization of AGN are greater and more variable at millimeter
than at centimeter wavelengths (e.g., Nartallo \etal 1998).

A polarization fraction less than 1\% is uncommon in compact radio
sources at centimeter wavelengths (Aller, Aller \& Hughes 1992).  However,
optically thick quasar cores observed with VLBI are frequently weakly polarized
(Cawthorne et al. 1993).  Such cores may be analogous to the radio source
in Sgr A*, which, due to its low power, may not produce the strong shocks 
in the jet
that are the source for higher polarization regions in quasars.
Weak polarization is
more common in galaxies than quasars or blazars and it is also more
common in compact-double 
sources or sources with irregular morphologies (Aller, Aller \& Hughes 1992).
A notable source with a very low polarization fraction is the radio galaxy 
3C 84, which has a very irregular morphology.  If the radio to
millimeter spectrum of Sgr A* does arise in a jet, the low power of this jet 
or environmental effects in the Galactic Center region  
may limit the magnetic field order.
Alternatively, low energy electrons in the synchrotron 
environment of Sgr A* may depolarize the source.

\end{document}